\documentclass[aps,letterpaper,10pt,twocolumn,showpacs,superscriptaddress]{revtex4-1}
\usepackage[utf8]{inputenc}

\usepackage{amsmath}
\usepackage{amsfonts}
\usepackage{graphicx}
\graphicspath{{./figures/}}
\usepackage{braket}

\usepackage[colorlinks=true,bookmarks=false,citecolor=blue,urlcolor=blue,linkcolor=blue]{hyperref}
\usepackage{siunitx}

\usepackage{todonotes}

\begin{document}

\title{ Low-error encoder for time-bin and decoy states for quantum key distribution }
\def\devicename{{MacZac}}
\def\deviceacronym{{(\textit{Mac}h-\textit{Z}ehder-S\textit{a}gna\textit{c})}}

\author{Davide~Scalcon}
\author{Elisa~Bazzani}
\affiliation{Dipartimento di Ingegneria dell'Informazione, Universit\`a degli Studi di Padova, via Gradenigo 6B, IT-35131 Padova, Italy}

\author{Giuseppe~Vallone}
\affiliation{Dipartimento di Ingegneria dell'Informazione, Universit\`a degli Studi di Padova, via Gradenigo 6B, IT-35131 Padova, Italy}
\affiliation{Padua Quantum Technologies Research Center, Universit\`a degli Studi di Padova, via Gradenigo 6A, IT-35131 Padova, Italy}
\affiliation{Dipartimento di Fisica e Astronomia, Universit\`a di Padova, via Marzolo 8, IT-35131 Padova, Italy}

\author{Paolo~Villoresi}
\affiliation{Dipartimento di Ingegneria dell'Informazione, Universit\`a degli Studi di Padova, via Gradenigo 6B, IT-35131 Padova, Italy}
\affiliation{Padua Quantum Technologies Research Center, Universit\`a degli Studi di Padova, via Gradenigo 6A, IT-35131 Padova, Italy}

\author{Marco~Avesani}
\email[Corresponding author: ]{marco.avesani@unipd.it}
\affiliation{Dipartimento di Ingegneria dell'Informazione, Universit\`a degli Studi di Padova, via Gradenigo 6B, IT-35131 Padova, Italy}

\begin{abstract}
Time-bin encoding has been widely used for implementing quantum key distribution (QKD) on optical fiber channels due to its robustness with respect to drifts introduced by the optical fiber. However, due to the use of interferometric structures,  achieving stable and low intrinsic Quantum Bit Error rate (QBER) in time-bin systems can be challenging.
A key device for decoy-state prepare \& measure QKD is represented by the state encoder, that must generate low-error and stable states with different values of mean photon number. Here we propose the \devicename \deviceacronym, a time-bin encoder with ultra-low intrinsic QBER ($<2\times 10^{-5}$) and high stability. The device is based on nested Sagnac and Mach–Zehnder interferometers and uses a single phase modulator for both decoy and state preparation, greatly simplifying the optical setup. The encoder does not require any active compensation or feedback system and it  can be scaled for the generation of states with arbitrary dimension.
We experimentally realized and tested the device performances as a stand alone component and in a complete QKD experiments. Thanks to the capacity to combine extremely low QBER, high stability and experimental simplicity the proposed device can be used as a key building block for future high-performance, low-cost QKD systems. 
\end{abstract}

\maketitle

\section{Introduction}
\label{sec:introduction}
    The security of transmitted data is a critical point in our modern and interconnected society. Encrypted data are constantly under attack and the breakthrough of quantum computing poses a radical threat to the security of the modern classical cryptographic protocols~\cite{Mavroeidis2018}. In this scenario, unconditional security can be guaranteed only by taking advantage of the same quantum mechanics framework, so that Quantum Key Distribution (QKD) protocols stand out. QKD enables to share a secret key between two parties by exchanging quantum states~\cite{Pirandola2019rev}, typically encoded in light's degrees of freedom (DoFs). A common encoding exploits the polarization DoF, which has been used to demonstrate QKD in both free-space ~\cite{Liao2017_daylight,Avesani2021} and fiber-based links ~\cite{Wang2016, Xavier2009, Agnesi20opt,Avesani2022}. While for free-space links, there is no need to compensate for polarization's changes in the channel, because of the negligible birefringence of atmosphere, for fiber optical channels the polarization drifts, induced  by random environmental stresses, demand for an active compensation of the state transformation. 
    
    By being insensitive to birefringence, time-bin has been extensively used as encoding in fiber optical QKD links as it overcomes the issue and removes the necessity of supervisioning perturbations in the channel. Reliability and robustness of time-bin encoding has been demonstrated by a number of experimental trials, in both laboratory environment~\cite{Liu2013, Yin2016, Boaron2018_GHz} and deployed fiber links~\cite{Tang2015, Tang2023}, with a plenty of different approaches mostly relying on fiber optics~\cite{Boaron2018,Boaron2018_GHz} or integrated photonics~\cite{Sibson:17,Sax2023}.

    However, due to the need to stabilize and retrieve the relative phase between the time-bins, time-bin encoding usually requires the stabilization of local interferometers. Implementations are typically characterized by a higher intrinsic QBER compared to polarization implementation~\cite{Agnesi20opt}.\par 
  
    One of the most widely used QKD protocols, implemented with both polarization and time-bin, is BB84~\cite{Bennett2014_BB84}. This protocol has undergone successive refinements over the years to enhance key rates and to adapt the protocol assumptions to accommodate technological limitations.
    As a matter of fact, the original BB84 protocol requires the use of four prepared states, but three states have been shown to be sufficient to ensure security without penalty on the key rate \cite{PhysRevA.97.042347,Rusca2018}, allowing a further simplification of the transmitter design.\par In addition, also the original requirement of using single photons has been relaxed.
    
    Photonic states can indeed be encoded in phase-randomized attenuated laser pulses, which, however, expose the BB84 protocol to Photon Number Splitting (PNS) attacks~\cite{Ltkenhaus2002}. A common countermeasure is the decoy state method~\cite{Ma2005}, which requires controlling the mean photon number of the transmitted pulse. For practical transmitters, this leads to the implementation of an intensity control stage, which is, in general, independent of the state encoder~\cite{Avesani:21}.\par 
   
    The present work proposes an efficient and simple encoder design for the preparation of both time-bin states of arbitrary dimensions and arbitrary decoy levels in a compact and single optical topology. 
    The device is all-fiber-based and exploits a Mach-Zehnder interferometer nested in a Sagnac loop, hence the name \devicename, to provide high extinction ratios and long-term stability. The \devicename can use as low as one phase modulator for both state encoding and decoy generation, greatly simplifying the experimental optical setup.
    We built an all-fiber prototype using only Commercial-Off-The-Shelves (COTS) components to characterize and verify the validity of the scheme for two-dimensional states (qubits). Thanks to its peculiar geometry, the device was capable of performing state encoding with an extremely high extinction ratio, showing a record-low intrinsic QBER of $2\times 10^{-5}$. The \devicename was then integrated into a complete QKD experiment to showcase its performance and advantages.
    By running the efficient three-state BB84~\cite{Fung2006}, with one-decoy method~\cite{Rusca2018_APL}, we achieved a steady distillation of a secret key, following from the extremely low QBER ($<0.03\%$ in key basis) provided by the encoder.\par
    Such device, combining high-performance, long-term stability with a practical and simple implementation, greatly simplifies the implementation of performant QKD transmitters, paving the way for more practical and low cost QKD systems. 
    
\section{Proposed scheme}
\label{sec:proposed_scheme}
    A common approach employed to generate time-bin states in QKD exploits an asymmetric interferometer, usually a Mach–Zehnder or Michelson type, which converts a pulsed laser source into a train of $N$ successive pulses with known temporal delay and defined phases $\phi_j$~\cite{Boaron2018_GHz, Mo2005, zhan2022}. The parameters $\phi_j$ are controlled by means of phase modulators, electro-optical devices that change their refractive index when driven by an external electrical signal.

    \begin{figure}[t]
        \centering
        \includegraphics[width=\columnwidth]{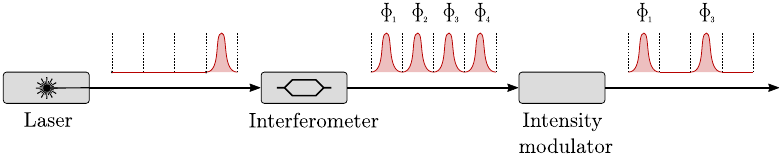}
        \caption{Typical approach for the generation of time-bin states in QKD experiments. Laser pulses are distributed over $N$ time bins (four in figure) and their phase is determined by an unbalanced interferometer structure. The desired amplitudes are selected by an intensity modulator.}
        \label{fig:block_diagram}
    \end{figure}
    
    The appropriate mean photon number of each pulse, as required by the decoy protocol, is then prepared by an intensity modulator, as sketched in Figure \ref{fig:block_diagram}, as sketched in Figure \ref{fig:block_diagram}.
    
    Multiple technologies are available to perform intensity modulation of laser light. Examples are electro-optic or electro-absorption devices, whose working principle relies on the controllable change in a material property to directly apply the intended modulation~\cite{Sinatkas2021}. The most common approach is to employ symmetric Mach–Zehnder inteferometers, which are available as COTS components and can support modulation bandwidths of several tens of \si{\giga\hertz}. However, these devices are sensitive to thermal and mechanical fluctuations and they require an addition stabilization mechanism to operate in a stable way.\par
    A simpler and more stable solution exploits unbalanced Sagnac interferometers, which do not require an active control since thermal and mechanical induced fluctuations are compensated by the double-pass topology ~\cite{Roberts2018}. Such scheme does not only provides a better long-term stability, but, at the same time, it greatly simplifies the experimental implementation. 
    
    \begin{figure*}[t]
        \centering
        \includegraphics[width=0.75\textwidth]{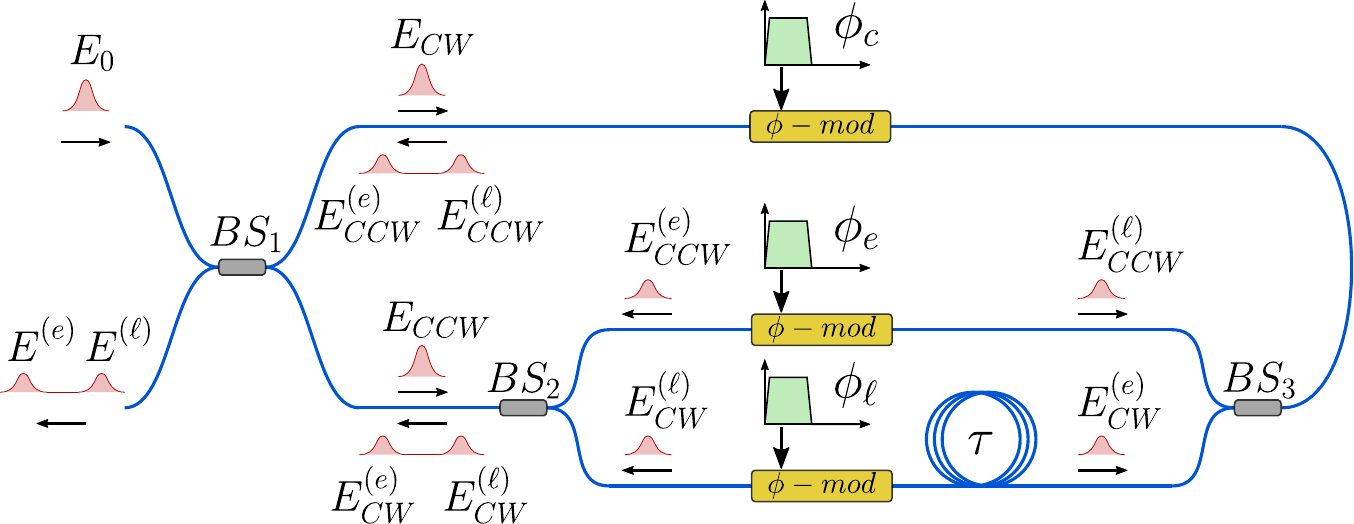}
        \caption{Proposed scheme for two-dimensional states generation. In this general scheme, each phase modulator is individually driven to modulate optical pulses traveling in a single direction: the phase modulator in the Sagnac arm modulates the $CW$ pulse, while the phase modulators in the Mach–Zehnder modulate the $CCW$ pulses.   BS: Beamsplitter, $\phi$-mod: phase modulator.} 
        \label{fig:sketch_seppia}
    \end{figure*}
    
    Based on the above observation, we here propose a scheme for the generation of time-bin states based on a unbalanced Mach–Zehnder interferometer included into a Sagnac based intensity modulator. The configuration allows to create time-bin states with extremely high-extinction ratio and decoy states with arbitrary intensity values, with a simple and stable setup. Since the loop layout avoids the use of an active compensation for relative phase drifts as mentioned before, the presented encoder ensures a practical and cheap implementation, where intensity and information encoding can be achieved by a single modulation stage, and besides, the design can be scaled to provide time states of arbitrary dimension.
    While the scheme can be employed for the generation of qu$d$it states (i.e. $d$ dimensional time-bin), as explained at the end of Sec. \ref{sec:expermental_results}, the following study starts by describing its working principles for qubits ($2$-dimensional states).

    In Figure \ref{fig:sketch_seppia} we show the design for qubits. Each 50/50 beam splitter (BS$_j$) is characterized by transmissivity and reflectivity coefficients $T_j=R_j=1/2$. 
    The input BS$_1$ equally divides the injected pulse into two modes that travel the Sagnac loop in clockwise (CW) and counterclockwise (CCW) directions, respectively. If $E_0$ is the complex amplitude at the input, the amplitude of the mode propagating in the CW (CCW) direction is expressed by Eq. \ref{eq:amplitudes_just_after_the_input_BS}:
    \begin{subequations}
    \label{eq:amplitudes_just_after_the_input_BS} 
        \begin{align}
            E_{CW} &= \frac{1}{\sqrt{2}} E_0 \,,\quad        E_{CCW} = i\frac{1}{\sqrt{2}}E_0 
        \end{align}
    \end{subequations}
    
    The CW mode ($E_{CW}$) first propagates in the phase modulator, that applies a phase shift $\phi_c$, and then it is transformed by the Mach–Zehnder interferometer, enclosed by BS$_2$ and BS$_3$, into a pair of pulses  - early ($e$) and late ($\ell$) - with identical amplitudes and with relative delay $\tau$. The other one ($E_{CCW}$) instead is first transformed into a pair of pulses, whose phases ($\phi_e$ and $\phi_\ell$) can be selected by the phase modulators in the Mach–Zehnder interferometer arms.
    Here, for the sake of simplicity, we are assuming that the phase modulator(s) that acts on the CW (CCW) mode, is turned off before the CCW (CW) mode light propagates through it, so that it doesn't apply any phase shift to the CCW (CW) propagating light. In other words, $\phi_c$ is applied only to the CW mode, while $\phi_{e,\ell}$ is applied only to the pair of CCW modes.\par
    Before being recombined by the input BS$_1$, the light pulses travelling in the loop have the complex amplitudes given by $E_{CW}=E_{CW}^{(e)}+E_{CW}^{(\ell)}$ (and similarly for $E_{CCW}$), where

    \begin{subequations}
    \label{eq:amplitudes_before_recombination_at_the_BS}
        \begin{align}
            E_{CW}^{(e)}  &= \frac{E_0}{2\sqrt2} e^{i\phi_c}\,,\quad
            &E_{CW}^{(\ell)}&= -\frac{E_0}{2\sqrt2} e^{i\phi_c}\\
            E_{CCW}^{(e)} &= i\frac{E_0}{2\sqrt2} e^{i\phi_e} \,,\quad
            &E_{CCW}^{(\ell)}  &= -i\frac{E_0}{2\sqrt2} e^{i\phi_\ell} 
        \end{align}
    \end{subequations}
    and we assumed for simplicity $e^{i\omega \tau}=1$, i.e. the relative phase imposed by the two arms of the unbalanced interferometer is zero. After the interference at the input BS$_1$, it follows that the complex amplitudes at the output of the device of the $e$ and $\ell$ pulses have the form of Eq. \ref{eq:early_output_amplitude}.

    \begin{subequations}
    \label{eq:early_output_amplitude}
        \begin{align}
            E^{(e)} 
            &= \frac{E_{CW}^{(e)} + i E_{CCW}^{(e)}}{\sqrt2}
            =\frac{i E_0 }{2}
            e^{i\phi^{e}_+}\sin\phi^{e}_-\\
            E^{(\ell)} 
            &= \frac{ E_{CW}^{(\ell)} + i E_{CCW}^{(\ell)}}{\sqrt2}
            =\frac{-iE_0 }{2}
            e^{i\phi^{\ell}_+}\sin\phi^{\ell}_-\,,\qquad \\
            &\mathrm{where} \qquad \phi^{e,\ell}_\pm = \frac{\phi_c \pm \phi_{e,\ell}}{2}
        \end{align}
    \end{subequations}

    {The output (unnormalized) state can be written as }
    \begin{equation}
        \ket{\psi_{\rm out}}=\sqrt{T_e}\ket{E}-e^{i\delta\phi}\sqrt{T_\ell}\ket{L}
    \end{equation}
    where $\ket{E}$ and $\ket{L}$ are the $e$ and $\ell$ time-bins while
    \begin{align}
    \label{eq:output_transmittance}
        T_{e,\ell} &= \frac{1}{4}\sin^2\frac{\phi_c-\phi_{e,\ell}}{2}\,,
        &e^{i\delta\phi}&=e^{i\frac{\phi_\ell-\phi_e}{2}}
    \end{align}

    From the above relation it is possible to see that the output quantum state can be controlled by properly setting the three phases $\phi_c$, $\phi_e$ and $\phi_\ell$.
    All BSs in Figure \ref{fig:sketch_seppia} are assumed to be ideal, with a splitting ratio exactly of $\frac{1}{2}$. 
    The analysis of the effect of imperfect BS is detailed in Appendix \ref{appendix:imperfections}, the impact of those differences is shown to be small.\par
    
\begin{figure}[t]
        \centering
        \includegraphics[width=\columnwidth]{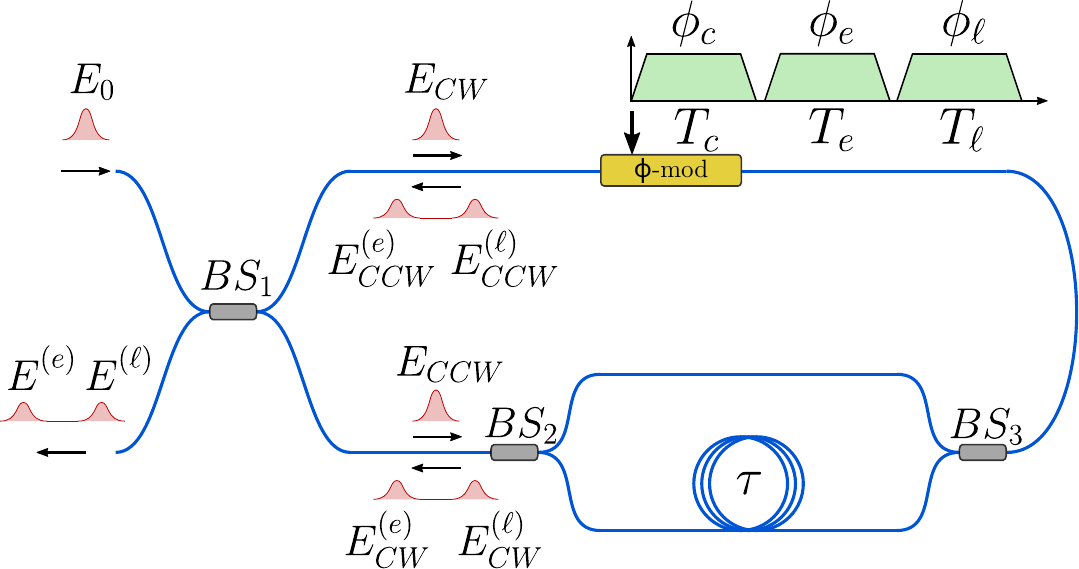}
        \caption{Simplified scheme for the two-dimensional state generation. In this scheme a single phase modulator is used and electrical pulses are sent at different times to modulate either the $CW$, the $CCW^{(e)}$ or the $CCW^{(l)}$ pulses. BS: Beamsplitter, $\phi$-mod: phase modulator.}
        \label{fig:sketch_seppia_single_phaseMod}
        
    \end{figure}

    In the analysis above, the phase modulators are assumed to apply a phase shift only to a single pulse of light propagating in a single direction.
    However, by considering a more general regime in which travelling wave effects~\cite{Li2004} are not relevant, those devices act as phase modulators on light propagating in both directions. Therefore, the scheme can be simplified by using one phase modulator (see Fig. \ref{fig:sketch_seppia_single_phaseMod}) as long as the different optical pulses $E_{CW}$, $E^{(e)}_{CCW}$ and $E^{(\ell)}_{CCW}$ travel the single modulator at distinct time instants. Indeed, in the latter condition, it is possible to apply arbitrary phases to the signals in the loop before they recombine. By suitably changing the modulating voltage signal at different times, the phase modulator can set the phases $\phi_c$, $\phi_e$, and $\phi_\ell$.
    
    In this scenario, to generate $\ket{E}$ ($\ket{L}$), the control electronics applies an electric pulse to the phase modulator when is travelled by the $e$ ($\ell$) component of the CCW light.
    The $\ket{-}$ state is produced by modulating the single CW laser pulse. To ensure the same mean photon number for the three states, the pulse applied to generate $\ket{-}$ has to be smaller than the one used for $\ket{E}$ and $\ket{L}$. Those amplitudes can be evaluated from the device response derived in Sec. \ref{sec:expermental_results}.
    Having a single phase modulator makes the overall device cheaper and simpler, but demands an higher modulation speed. The device must be able to switch between the required phase values in a time comparable with the distance between the pulses of the time-bin state. On the other hand, the bandwidth constraint from the modulators in Figure \ref{fig:sketch_seppia} is weaker, since the limiting factor arises from the repetition rate and the propagation delay between CW and CCW modes.

    The asymmetric Sagnac structure, which inherently implements an intensity modulator, allows to encode time-bin states and, at the same time, their decoy levels. 
    
    Since the transmittance is function of the applied phase shift, the mean photon number is set by controlling the amplitude of the signals driving the phase modulators. From the experimental point of view, this can be implemented with a digital to analog converter that is fast enough to generate the electrical signals driving the electro-optic phase modulators, as described in Appendix \ref{appendix:electronics}.

\section{Experimental results}
\label{sec:expermental_results}
    In this work we were interested in developing and evaluating the performance of the \devicename in its simplest form, with only one modulator, since this provides the highest benefits in terms of optical implementation and cost-reduction. In the following we describe its experimental implementation and characterization.

    \subsection{Transmitter characterization}
    \label{subsec:transmitter_characterization}
    
\begin{figure}
            \centering
            \includegraphics[width=0.8\columnwidth]{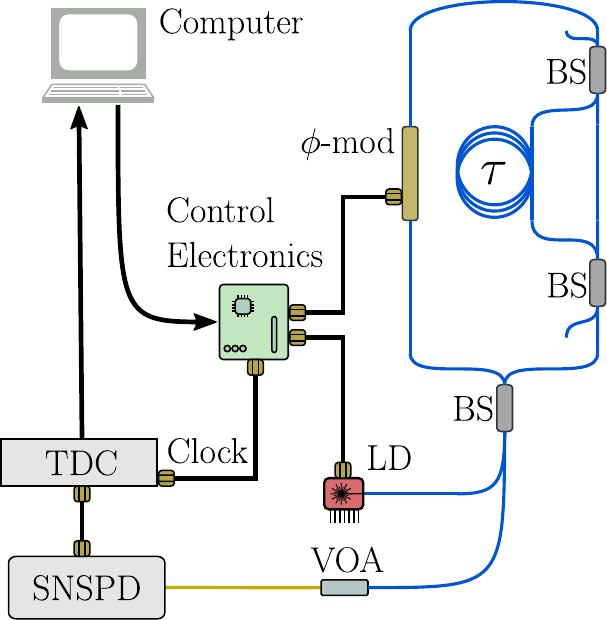}
            \caption{Source characterization setup. LD: laser diode, BS: beamsplitter, $\phi$-mod: phase modulator, TDL: tunable delay line, VOA: variable optical attenuator, TDC: time to digital converter, SNSPD: superconducting nanowire single photon detector.}
            \label{fig:characterization_setup}
        \end{figure}

        To measure the achievable performances of the encoder we built an experimental setup with COTS fiber components, as represented in Figure \ref{fig:characterization_setup}. 
        Specifically, the setup implements the single phase modulator layout (Fig. \ref{fig:sketch_seppia_single_phaseMod}) and it includes an unbalanced Mach-Zehnder inteferometer with a delay of about 10ns, corresponding to 2 meters of optical fiber. Shorter delays, and thus higher repetition rates are possible if the fiber components are shortened and spliced togheter.

        All fiber components are polarization maintaining, BSs are fast axis blocking, i.e. they act as polarizing elements suppressing the mode travelling along the fast axis, and the phase modulator is a Lithium Niobate fiber modulator with $10$\si{\giga\hertz} electro-optical bandwith. 
        The light source is composed by a 1550nm distributed feedback laser diode, driven in gain-switch mode, which generates laser pulses of about 50ps width FWHM at 10MHz repetition rate.
        
        \begin{figure*}
            \centering
            \includegraphics[width=\textwidth]{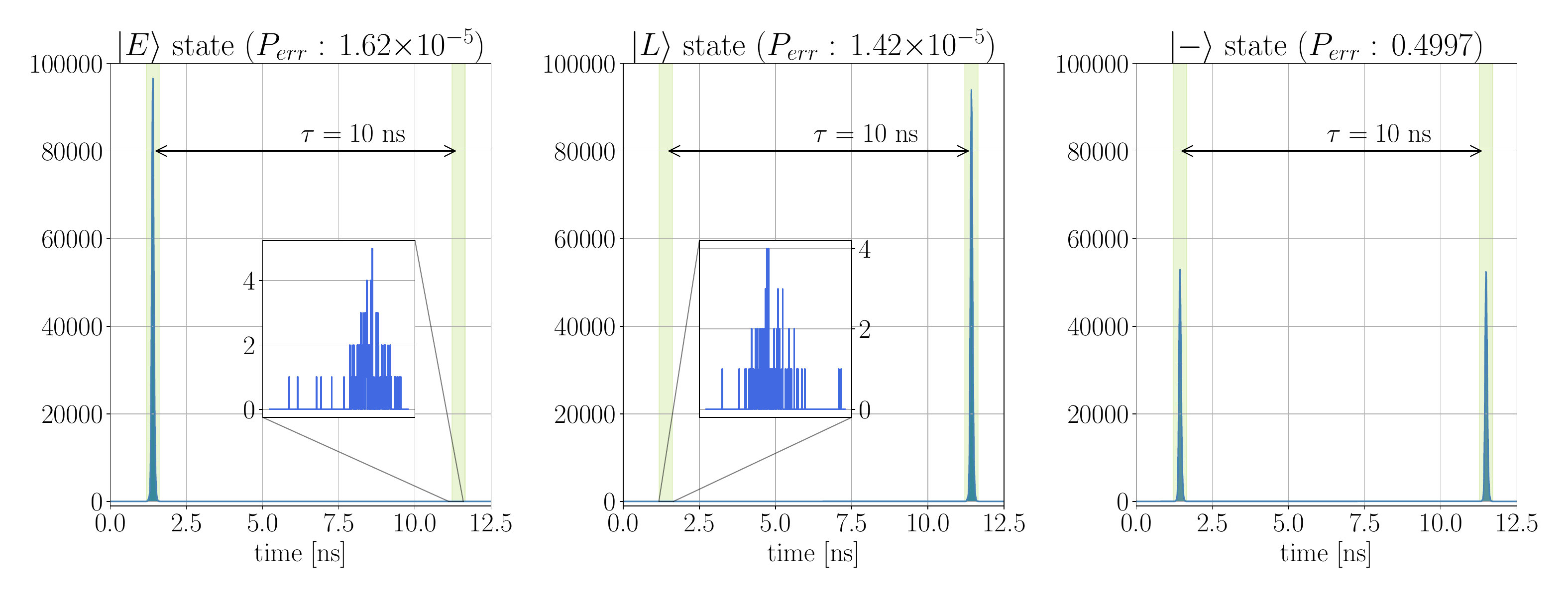}
            \caption{Source characterization results, for early (E, left), late (L, center) and the superposition (+, right) states. The diagrams represent the histogram of the time of arrival of the photons produced by the source and the corresponding extinction ratio.}
            \label{fig:QBER_characterization}
        \end{figure*}
        
        The control electronics (for details see Appendix \ref{appendix:electronics}) is designed to generate three states: $\ket{E}$, $\ket{L}$ and the superposition $\ket{-} = \frac{1}{\sqrt{2}}\left( \ket{E} - \ket{L}\right)$, each with two possible amplitude levels $\mu$ and $\nu$, with $\nu < \mu$. This enables to implement the three-states one-decoy efficient BB84 protocol~\cite{Grunenfelder2018}.

        The states $\ket E$ ($\ket L$) with amplitude $\mu$ are obtained by setting phases $\phi_\ell = \phi_c = 0$ ($\phi_e = \phi_c = 0$) and $\phi_e = \alpha_\mu$ ($\phi_\ell = \alpha_\mu$) with arbitrary $0\leq\alpha_\mu\leq\pi$. On the other hand, to obtain the amplitude $\nu$, the applied phase is $\alpha_\nu$ such that
        
        \begin{equation}
            \sin^2\left(\frac{\alpha_\nu}{2}\right)=\frac{\nu}{\mu}\sin^2\left(\frac{\alpha_\mu}{2}\right)
        \end{equation}
        We note that the above equation define the values of $\alpha_\nu$ in function of any arbitrary values of $\alpha_\mu$ as long as $\nu<\mu$.

        The state $\ket -$ is generated by setting $\phi_e = \phi_\ell = 0$. The value of $\phi_c$ is set to $\beta_\mu$ for the high-intensity signal level $\mu$ and it is set to $\beta_\nu$ for the lower amplitude signal. The relation between $\beta_\mu$ and $\beta_\nu$ is the same as the relation between $\alpha_\mu$ and $\alpha_\nu$ while, to guarantee the same mean photon number in the produced $\ket E$ or $\ket L$ and $\ket -$ states, the following relations between $\alpha_{\mu,\nu}$ and $\beta_{\mu,\nu}$ must hold
        \begin{align}
            \sin^2\left(\frac{\beta_{\mu,\nu}}{2}\right)=\frac{1}{2}\sin^2\left(\frac{\alpha_{\mu,\nu}}{2}\right) 
        \end{align}
        The applied phases for each generated state is summarized in Table \ref{tab:electrical_signals_table}.
        The control electronics drives the phase modulator in three different time slots $T_E$, $T_L$ and $T_-$.
        The first two are the instants in which the CCW pulses cross the phase modulator and, whenever $\phi_e$ and $\phi_{\ell}$ are applied, they correspond to the generation of $\ket{E}$ and $\ket{L}$ respectively. The latter allows to produce $\ket{-}$ by setting $\phi_c$ to the single CW pulse.

        \begin{table}
        \centering
        \caption{Amplitude and time position of the pulses generated by the control electronics for all possible state to be prepared. Time instant $T_E$ ($T_L$) means that the output is triggered when the early (late) component of the counterclockwise signal crosses the phase modulator, $T_+$ specifies that the output instant coincides with the time in which the clockwise pulse goes through the phase modulator.}
        \label{tab:electrical_signals_table}
            \begin{ruledtabular}
                \begin{tabular}{c l | c c c c}
                    State & Decoy & Time slot &$\phi_c$ & $\phi_e$ & $\phi_\ell$  \\ \hline
                    $\ket{E}$ & $\mu$ (decoy high) & $T_E$ & $0$ & $\alpha_\mu$ & $0$ \\
                    $\ket{E}$ & $\nu$ (decoy low)  & $T_E$ & $0$ & $\alpha_\nu$ & $0$ \\
                    $\ket{L}$ & $\mu$ (decoy high) & $T_L$ & $0$ & $0$ & $\alpha_\mu$ \\
                    $\ket{L}$ & $\nu$ (decoy low)  & $T_L$ & $0$ & $0$ & $\alpha_\nu$ \\
                    $\ket{-}$ & $\mu$ (decoy high) & $T_-$ & $\beta_\mu$ & $0$ & $0$ \\
                    $\ket{-}$ & $\nu$ (decoy low)  & $T_-$ & $\beta_\nu$ & $0$ & $0$ \\
                \end{tabular}
            \end{ruledtabular}
        \end{table}

        To evaluate the quality of the source, we tested it by preparing a stream of identical states which, after being attenuated down to less than one photon per pulse with a variable optical attenuator, were measured by means of a SNSPD (superconducting nanowire single photon detector). Detection events were collected by a time tagger, whose clock was synchronized with the one of the control electronics. 
        Figure \ref{fig:QBER_characterization} shows the histogram of the time of arrival over a period. These data are used to measure the intrinsic QBER $P_{err}$ of the source, i.e.
        \begin{equation}
            P_{err} =\frac{N_{\rm wrong}}{N_{\rm wrong}+N_{\rm correct}}
            \label{eq:Perr}
        \end{equation}
    
        where $N_{\rm correct}$ are the number of photons detected in the correct time slot, and $N_{\rm wrong}$ the events happened in the wrong one.An other useful and closely related metric is the Extinction Ratio (ER) that, expressed in dB, can be evaluated from $P_{err}$ by taking $10\log_{10}\frac{N_{correct}}{N_{wrong}} = 10\log_{10}(\frac1{P_{err}}-1)$.
        
        By acquiring data for one minute, with a detection rate of about 140k detections per second, we measured $P_{err} = 1.62\times10^{-5}\pm1.4\times10^{-6}$ ($\mathrm{ER}=47.9$$\pm$0.4dB) for the early state and $P_{err} = 1.42\times10^{-5}\pm1.3\times10^{-6}$ ($\mathrm{ER}=48.5$$\pm$0.4dB) for the late state. For the superposition state, which requires the same number of events in both time bins, we measured $P_{err}=0.4997\pm2\times10^{-4}$. 
        Measurement errors are ascribed to the resolution ($\sim1$ps) and jitter ($\sim10$ps) of the timetagger and the jitter ($<30$ps) and darkcounts ($\sim200$Hz) of SNSPDs.
    
        The intrinsic $P_{err}$ of the encoder shows unprecedented low values compared to the literature. In some recent QKD demonstrations, the sources were indeed able to achieve an intrinsic QBER $P_{err}$ as low as $\sim10^{-3}$ \cite{Boaron2018_GHz, Roberts2017}, which is two order of magnitude greater than the result presented by this work.
        
        \begin{figure*}
        \includegraphics[width=\textwidth]{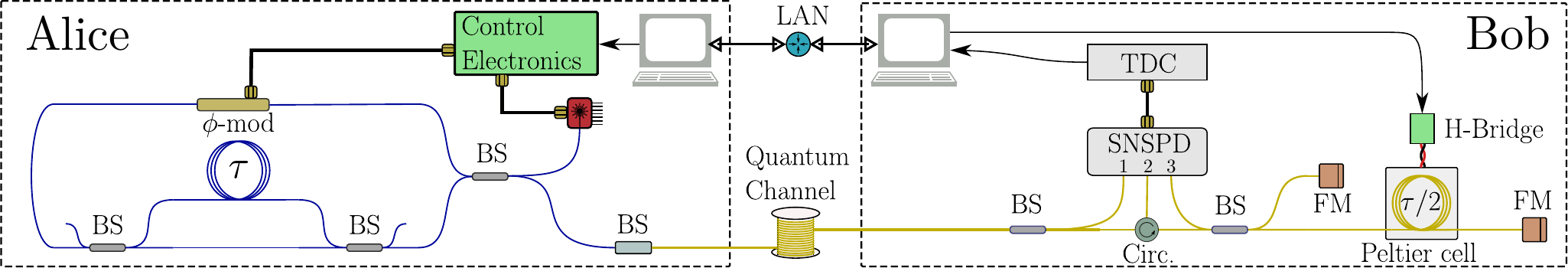}
            \caption{Experimental setup of the QKD experiment using the proposed encoder. LD: laser diode, BS: beamsplitter, $\phi$-mod: phase modulator, TDL: tunable delay line, VOA: variable optical attenuator, Circ: circulator, FM: Faraday mirror, TDC: time to digital converter, SNSPDs: superconducting nanowire single photon detectors.}
            \label{fig:QKD_setup}
        \end{figure*}

    \subsection{QKD experiment}
    \label{subsec:QKD_experiment}

        We used the transmitter characterized in Sec. \ref{subsec:transmitter_characterization} to carry out a complete QKD experiment. The electronic control is based on a FPGA (Field Programmable Gate Array) which provides the interface between the computer and the other electronic components, including the digital to analog converter described in Appendix \ref{appendix:electronics} and the laser driver. The FPGA design is an upgrade of the stack presented \cite{Stanco2022}, to accommodate the different hardware.
        
        The full setup encompassed an all-fiber receiver, based on the Michelson-Faraday interferometer~\cite{Mo2005}, and SNSPDs as detection system.  The optical channel consisted of a spool of single mode fiber with a length of 50km (0.2 dB/km). Due to the additional contributions of fiber connectors and detector inefficiency, the overall losses were estimated at about 16.4dB.The experimental setup is represented in Figure \ref{fig:QKD_setup}.\par
        
        To compensate for the relative phase drift between the transmitter and receiver interferometers, we implemented a control system based on a Peltier cell, acting on the receiver. The fiber's delay line loop at the receiver is mounted on one of the faces of the Peltier. The device slightly changes the temperature of one of the receiver's interferometer arm, changing the refractive index of the optical fiber and, thus, changing the phase. The receiver side computer controls a H-bridge circuit that delivers the appropriate current to the Peltier cell. The control logic is accomplished by a gradient descent algorithm which aims to maximize the measured ER of the control state $\ket-$, by applying a proportional and derivative correction to the Peltier current.
        
        The entire optical receiver was enclosed in a box to provide the best possible thermal insulation from the environment.\par 
        Temporal reference sharing and synchronizaion is performed by Qubit4sync algorithm~\cite{Calderaro2020}, which employes the same qubits for synchronizing the devices, avoiding the use of an additional, dedicated hardware. Bob recovers indeed the signal period directly from the arrival times of Alice's qubits, while the initial delay is determined by cross-correlating Bob's signal with a synchronization string, sent by Alice at the beginning of the transmission.
        
        The real-time QKD postprocessing was achieved by using an adapted version of the AIT QKD R10 software suite~\cite{AIT}, revised according to the finite-key analysis in~\cite{Rusca2018}, which gives the secret key length $l$ 

        \begin{subequations}
        \label{eq:SKR_rusca}
            \begin{align}
                l &\leq s_{Z,0}^l + s_{Z,1}^l(1- h(\phi_{Z}^u)) - \lambda_{EC} + \nonumber\\
                  &- 6\mathrm{log}_2(19/\epsilon_{sec}) - \mathrm{log}_2(2/\epsilon_{corr})
            \end{align}
        \end{subequations}
        where $s_{Z,0}^l$ and $s_{Z,1}^l$ are the lower bounds respectively on the vacuum and single-photon detection events, $h(\cdot)$ is the binary entropy function, $\phi_{Z}^u$ is the upper bound on the phase error rate, $\lambda_{EC}$ is the number of bits revealed during the error correction step and $\epsilon_{sec}$ and $\epsilon_{corr}$ are the secrecy and correctness parameter, respectively.\par
        
\begin{figure}[h]
            \centering
            \includegraphics[width=\columnwidth]{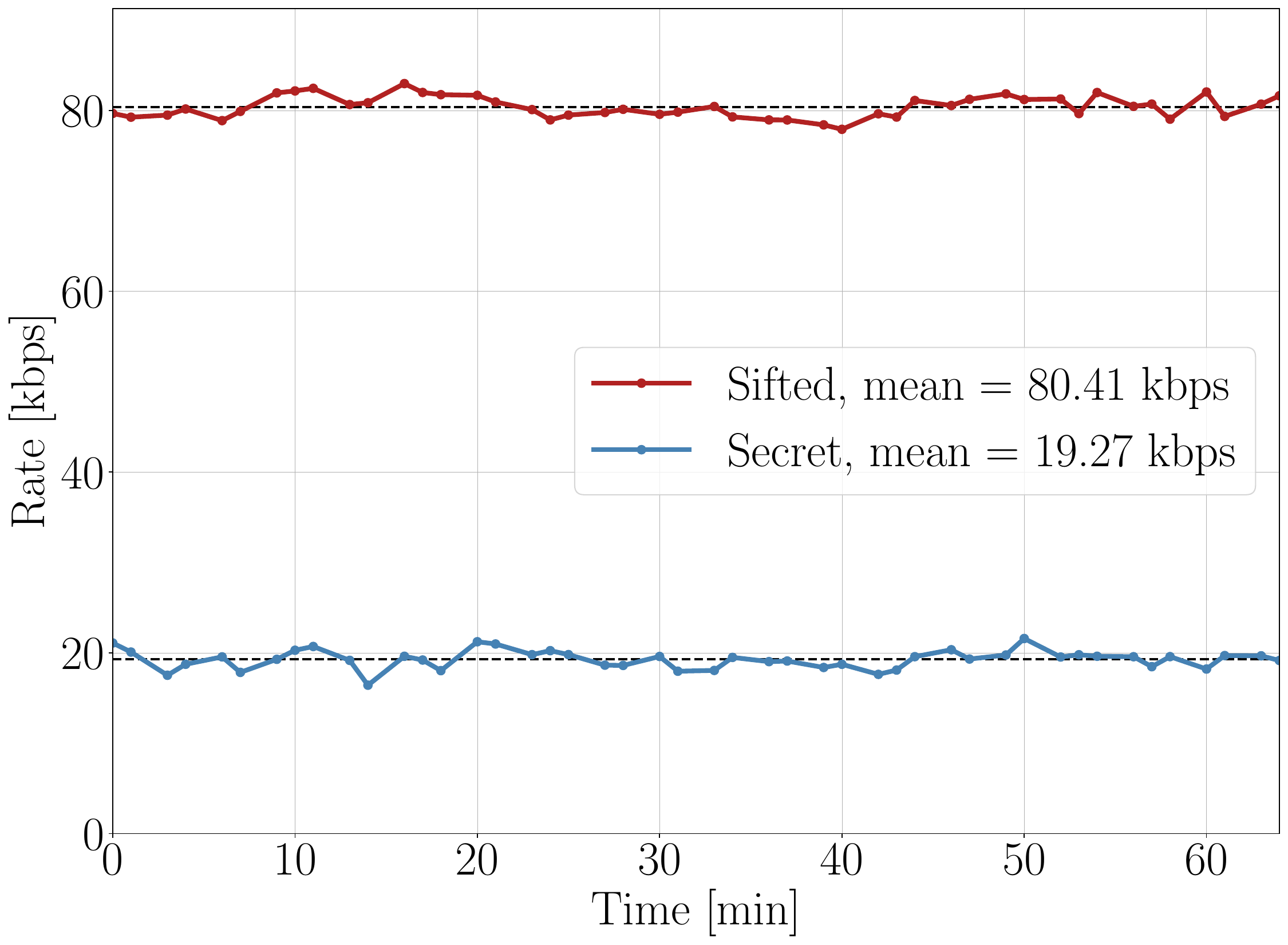}
            \caption{Secret key rate (blue) and sifted key (red) in the experiment.}
            \label{fig:SKR_over_time}
        \end{figure}

\begin{figure}[h]
            \centering
            \includegraphics[width=\columnwidth]{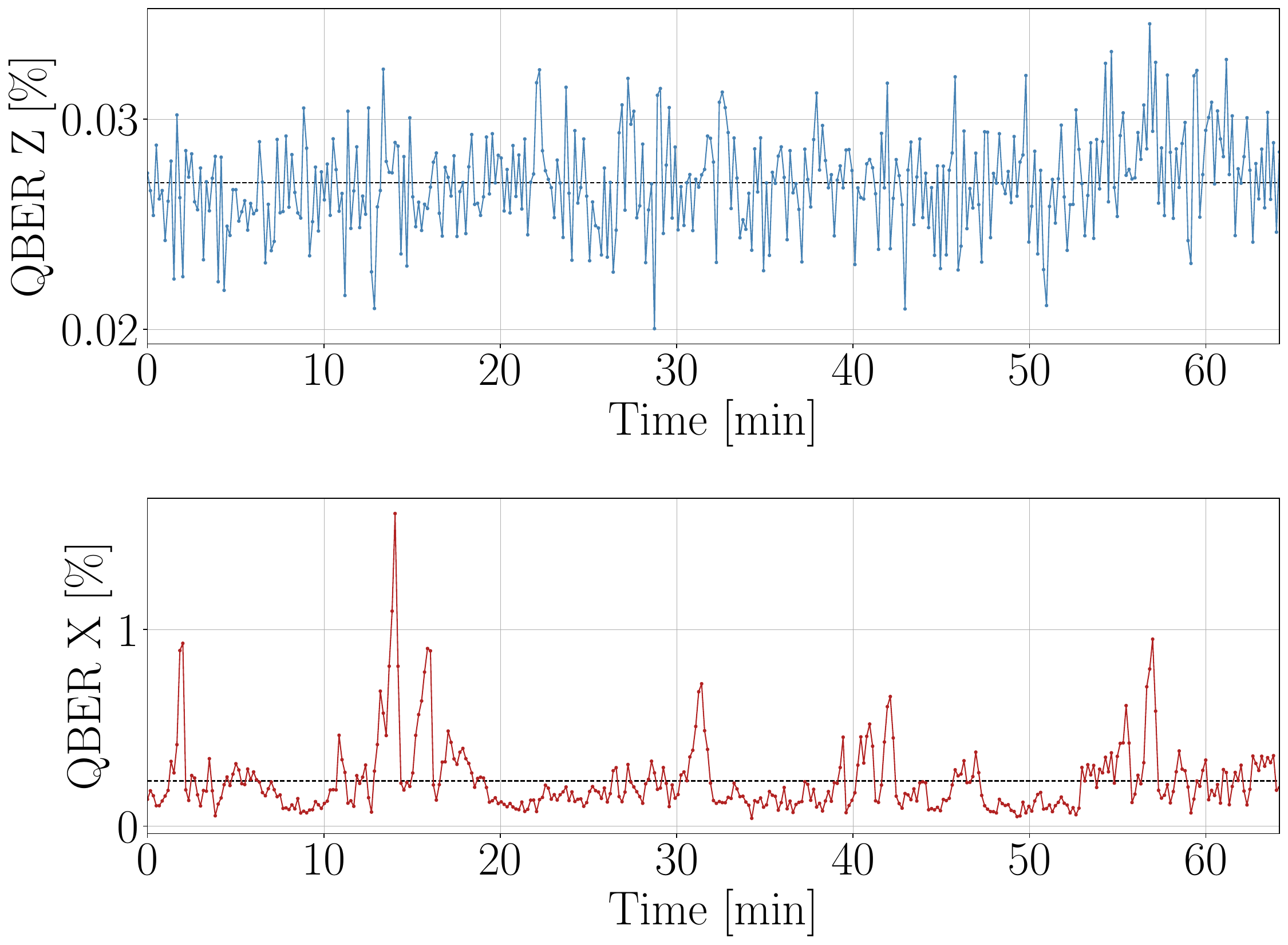}
            \caption{QBER in the $\mathcal{Z}$ basis (blue) and in the $\mathcal{X}$ basis (red) in the experiment.}
            \label{fig:QBER_over_time}
        \end{figure}

\begin{figure}[h]
            \centering
            \includegraphics[width=\columnwidth]{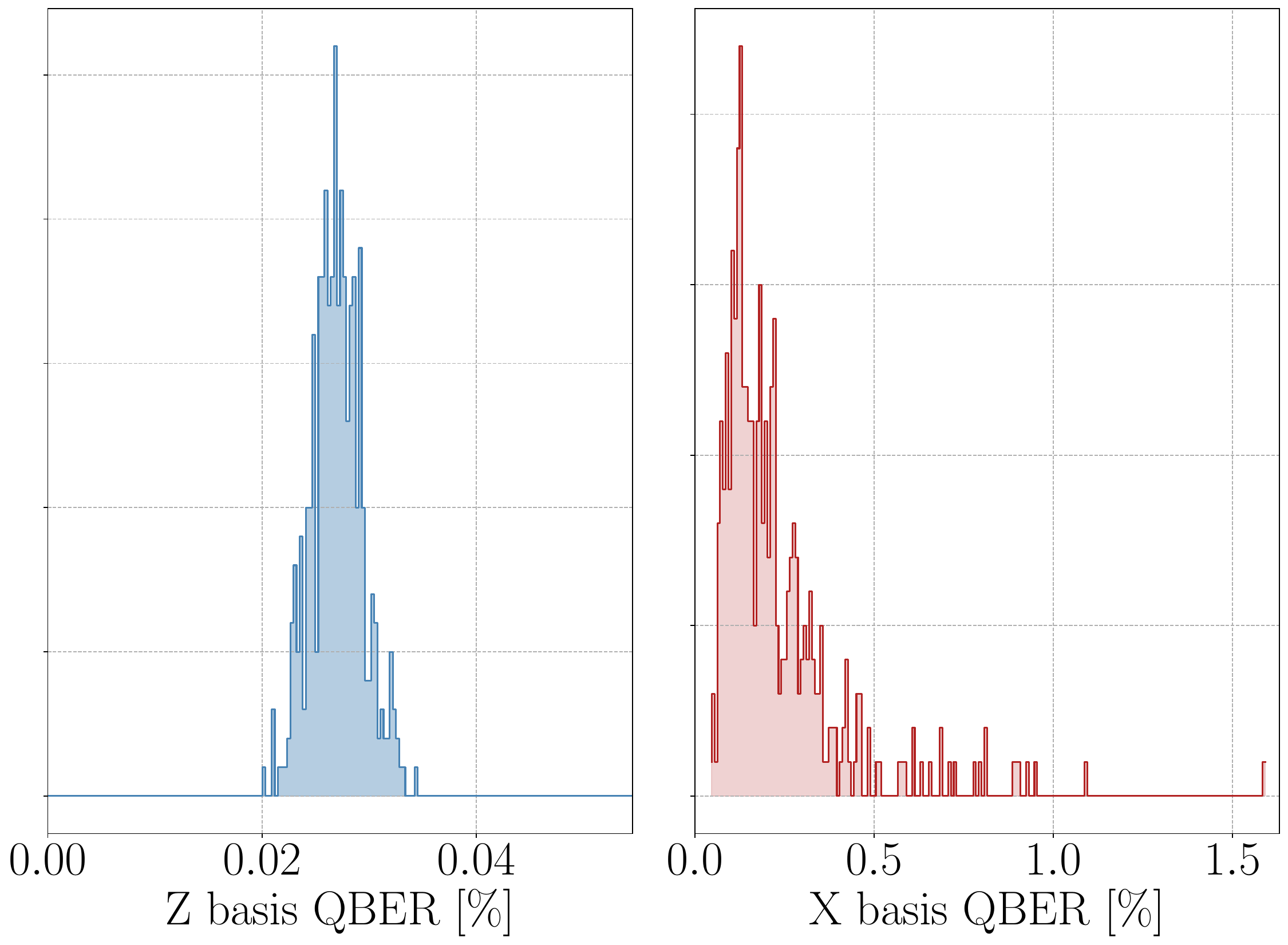}
            \caption{Distribution of the QBER in the $\mathcal{Z}$ basis (blue) and in the $\mathcal{X}$ basis (red) in the experiment. }
            \label{fig:QBER_distribution}
        \end{figure}

        We report the QBER and the sifted and secret key rate (SKR) achieved from 1 hour of run in Figures \ref{fig:QBER_over_time} and \ref{fig:SKR_over_time} respectively, while Tab. \ref{tab:QKD_experiment_results} summarizes the average values of the live demo, including also the detection rate. QBER in the $\mathcal{Z}$ basis is estimated on average as low as $0.027\%$; moreover, with a standard deviation of $0.012\%$, it exhibits an excellent stability, remaining within one standard deviation from its mean for the entire acquisition.
        We emphasizes that the great consistency of the results is achieved without any active control on the encoder, which is replaced by the self-compensation provided by the Sagnac loop. The control basis shows one order of magnitude higher mean QBER, i.e.$0.23\pm 0.29\%$, anyway still below $1\%$ on average. This is due to the high sensitivity of interferometers to environmental changes and the slow response of the temperature-based control system, as highlighted by the sharp peaks in the QBER time evolution (Fig. \ref{fig:QBER_over_time}, down). The different behavior over time of the QBERs in the two bases is pointed out also by their distributions represented in Figure \ref{fig:QBER_distribution}. QBER $\mathcal{Z}$ is en fact almost normal distributed around its average, whereas QBER $\mathcal{X}$ distribution shows an asymmetric tail extending toward higher QBER values. Nevertheless, the intrinsic reliability of the encoder allowed to achieve a SKR $19.27$ kbps, with less than $8\%$ fluctuations. 
        
        The low QBER in the key basis, which is independent of the receiver interferometer, constitutes the most relevant achievement of this work, as it certifies the inherent capabilities of the \devicename. The corresponding values results more than one magnitude lower with respect to recent time-bin experiments~\cite{Scalcon2022}\cite{Bouchard2022}, also exploiting Sagnac-based design~\cite{Tang2023}, and integrated photonic solutions~\cite{Sax2023,Yoshino2007,Sibson:17,zhan2022}.       
        
        \begin{table}[h]
            \centering
            \caption{Averages and standard deviations of the main results of the QKD experiment.}
            \label{tab:QKD_experiment_results}
            \begin{ruledtabular}
                \begin{tabular}{l | c c }
                                            & mean      & std   \\ \colrule
                    QBER Z [\%]             & 0.027     & 0.012 \\
                    QBER X [\%]             & 0.23      & 0.29  \\
                    SKR [kbps]              & 19.3      & 1.5   \\
                    Sifted [kbps]           & 80.4      & 1.2   \\
                    Detection rate [kHz]    & 112.9     & 1.7 
                \end{tabular}
            \end{ruledtabular}
        \end{table}
        
\section{Qudit generalization}
\begin{figure}
        \centering
        \includegraphics[width=\columnwidth]{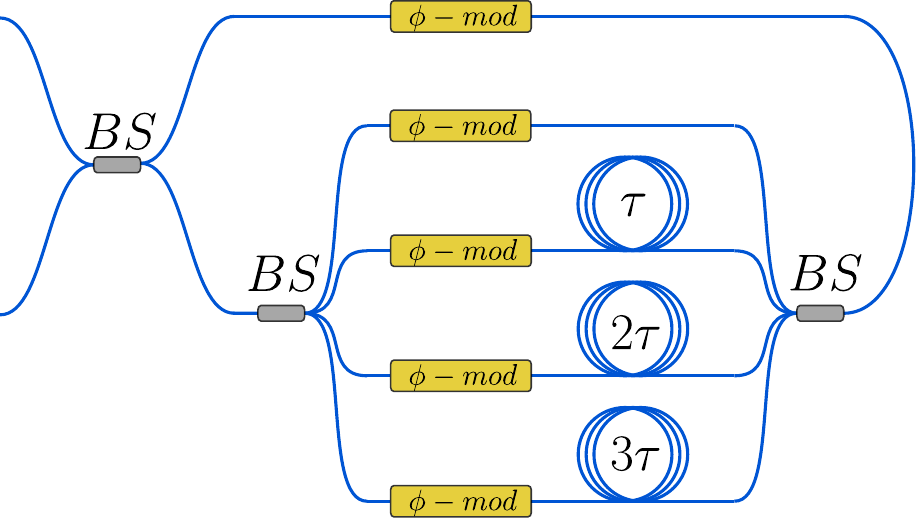}
        \caption{Proposed scheme for the generation of four dimensional time-bin states. BS: Beamsplitter, $\phi$-mod: phase modulator.}
        \label{fig:sketch_seppia_four_dim}
    \end{figure}

    The two-dimensional case described above can be generalized for the generation of $d$-dimensional states, by increasing the number of arms in the Mach-Zehnder interferometer to $d$. For instance, the design with $d = 4$ is shown in Figure \ref{fig:sketch_seppia_four_dim}.\par
    
    Assuming BS splitting as 50/50, the complex amplitude of the transmitted light in the $k-$th time slot can be written in the form of Eq. \ref{eq:nth_slot_amplitude}

    \begin{align}
        \label{eq:nth_slot_amplitude}
        E^{(k)} 
        &= \frac{iE_0}{d} e^{\phi^{(k)}_+} \sin{\phi^{(k)}_-} \,,\quad \phi^{(k)}_\pm = \frac{\phi_c\pm \phi_k}{2}
    \end{align}
    
    while the corresponding transmittance is given in Eq. \ref{eq:nth_slot_transmittance}
    \begin{align}
    \label{eq:nth_slot_transmittance}
        T^{(k)}_{\phi_c, \phi_k} = \frac{|E_{out}^{(N)}|^2}{|E_{in}|^2} = \frac{1}{d^2} \sin^2\left( \frac{\phi_c - \phi_k}{2} \right)
    \end{align}

\begin{figure}
        \centering
        \includegraphics[width=\columnwidth]{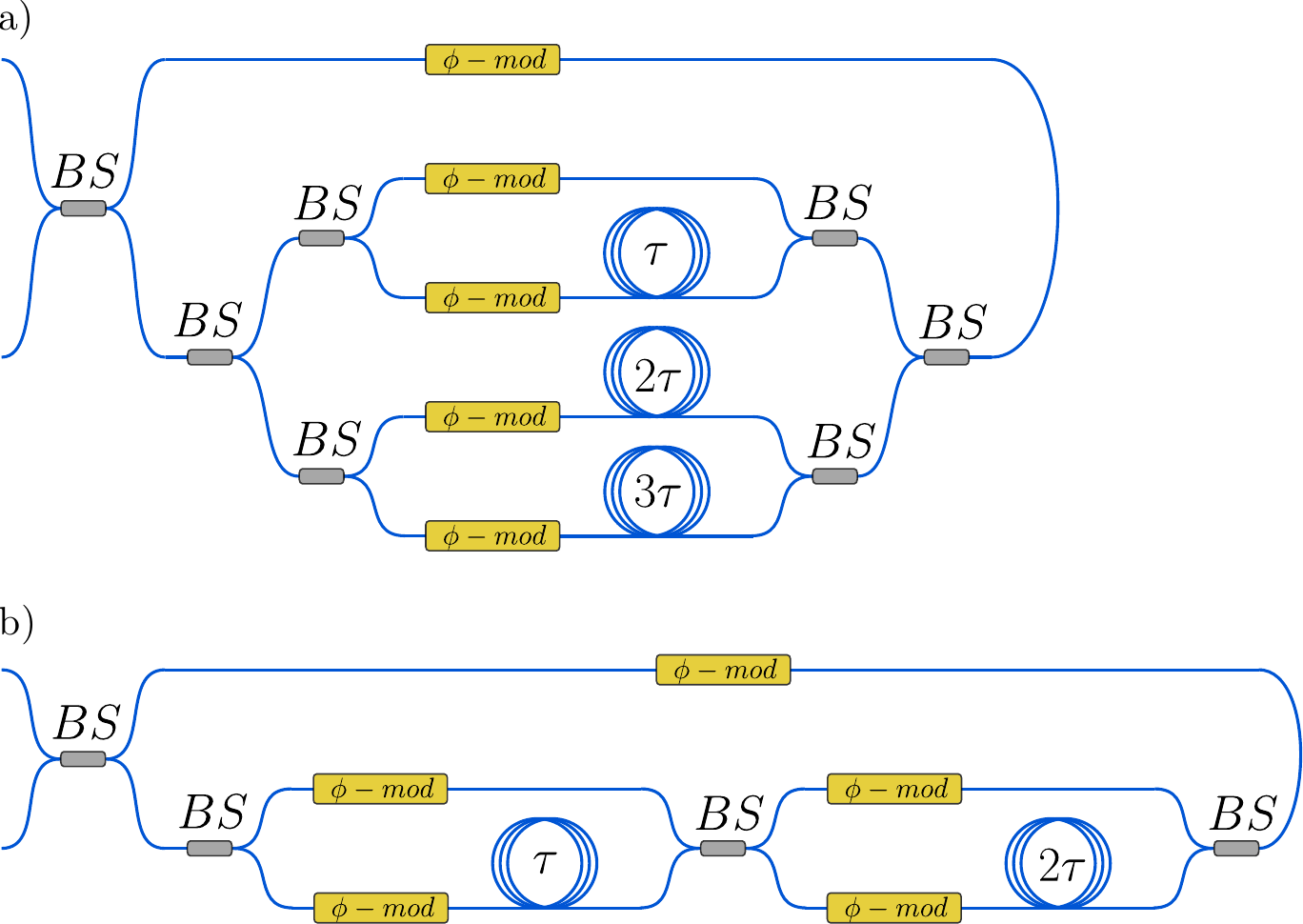}
        \caption{Scheme for the generation of four dimensional states using two by two beamsplitters. \textit{a)} Using a series of two interferometers, \textit{b)} Using two interferometers in parallel.  BS: Beamsplitter, $\phi$-mod: phase modulator.}
        \label{fig:four_dim_with_multi_BS}
    \end{figure}

    The $d$-arm interferometer requires BSs with at least d+1 ports, one input and $d$ outputs. If not available, the latter device can be replaced by a cascade of standard BSs, as shown in Figure \ref{fig:four_dim_with_multi_BS}.\par
    The main advantages of the proposed setups are the simplicity and scalability, as the structure can be extended to generate arbitrary dimension time-bin states. Moreover, the design allows to decrease to one the number of phase modulators, provided that the bandwidth of the single phase modulator is sufficient to act separately on the $d$ time bins. 

\section{Conclusions}
    In this work we propose an optical scheme, the \devicename, based on the Sagnac and Mach-Zehnder interferometers, for the generation of time-bin states of arbitrary dimensions and mean number of photons. 
    Such device significantly simplifies the transmitter design in a time-bin QKD experiment, by embedding in a single topology both the state preparation and the intensity modulation for the decoy state protocol.
    The capabilities of the proposed setup have been experimentally verified by means of an all-fiber system producing two-dimensional states. Remarkably, the intrinsic QBER of the encoder has been measured lower than $2\times10^{-5}$, without requiring for any compensation and initial calibration. This result is unprecedented, being the lowest reported in the literature~\cite{Boaron2018_GHz}~\cite{Roberts2017} to our knowledge.  
    The designed transmitter is applied to a QKD system to perform a real-time demonstration, in a laboratory environment, of the 3-states 1-decoy efficient BB84 protocol. On average, secret keys were generated at a rate of more than 19kbps with a preparation rate of 10MHz. The proposed transmitter demonstrated noteworthy performances in terms of key basis QBER, measured 0.027\% with a standard deviation of 0.012\%, direct consequence of the high stability and extinction of the encoder. 
    At the receiver side, the control basis was measured by an unbalanced Faraday-Michelson interferometer, in which the phase compensation is provided by a Peltier cell in order to minimize the receiver's losses. This method allowed us to achieve a low average QBER of 0.225\% for all the duration of the experiment.
    We think that this scheme might be useful in practical QKD apparatus, given the extremely low QBER, and the possibility to implement it with various technologies, such as discrete fiber components or a single, integrated, photonic chip, that guarantees stability and compactness.
    
    This work opens up to future investigations in the use of the same architecture, by increasing the dimensionality and the transmission rate, in order to achieve higher key rate for a practical secure communication. 
    
\section*{Acknowledgment}
    This work was partially supported by \textit{Quantum technologies 4 SECurity - Generazione quantistica di chiavi sicure (Q4Sec)} funded by Italian Defence Ministry within the National Military Research Plan (PNRM), and by European Union’s Horizon Europe research and innovation programme under the project {\it  Quantum Secure Networks Partnership (QSNP)}, grant agreement No 101114043. 
    Views and opinions expressed are however those of the author(s) only and do not necessarily reflect those of the European Union or European Commission-EU. Neither the European Union nor the granting authority can be held responsible for them.

    The device here presented is the object of the Italian Patent Application No. 102022000024354 filed on 25.11.2022\cite{Avesani2022_patent}.

\appendix
\newpage

\section{Control electronics}
\label{appendix:electronics}
\begin{figure}[ht]
        \centering
        \includegraphics[width=\columnwidth]{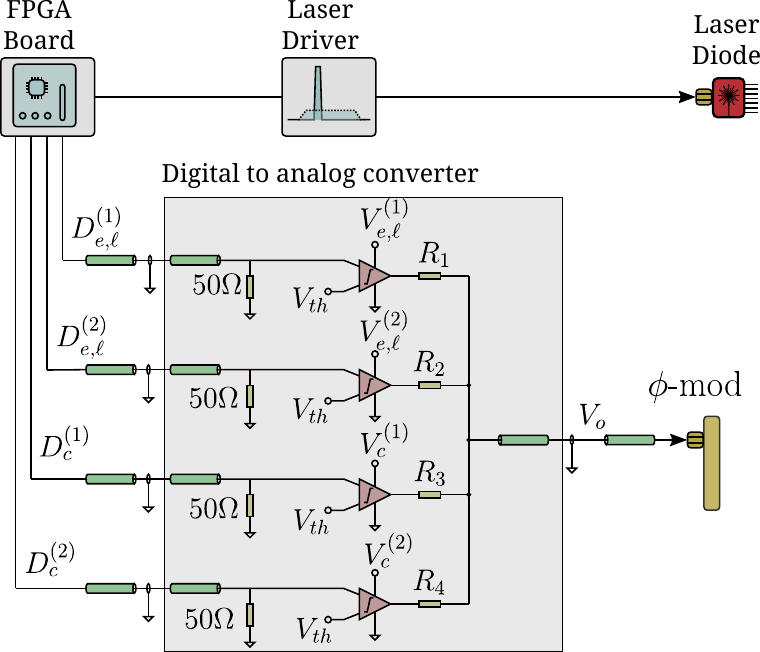}
        \caption{Electrical scheme of the used DAC.}
        \label{fig:electronics_block_diagram}
    \end{figure}

    The control electronics mentioned in Sec. \ref{subsec:transmitter_characterization} is composed by two parts: a commercial FPGA based board (ZedBoard by Avnet, based on the Zynq-7000 SoC) and a custom Digital to Analog Converter (DAC), developed in our lab and controlled by the FPGA. The block diagram is depicted in Figure \ref{fig:electronics_block_diagram}.
    
    \begin{figure}[ht]
        \centering
        \includegraphics[width=\columnwidth]{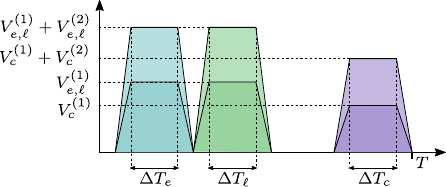}
        \caption{Graph of the signals generated by the digital to analog converter used to drive the phase modulator in the experiment.}
        \label{fig:Oscilloscope_traces1}
    \end{figure}
    
    We use four digital outputs of the FPGA, called here $D_{e,\ell}^{(1)}$, $D_{e,\ell}^{(2)}$, $D_{c}^{(1)}$ and $D_{c}^{(2)}$, the first two for the generation of $\ket{E}$ and $\ket{L}$ states, and the other two for the generation of the $\ket{-}$ state. To produce $\ket{E,\mu}$ (early with high decoy value) the FPGA generates a pulse at the time $T_E$, i.e. the instant in which early light travels through the phase modulator, on both $D_{e,\ell}^{(1)}$ and $D_{e,\ell}^{(2)}$ digital outputs. This signal triggers the two comparators $1$ and $2$, whose outputs are then added by the passive resistive network. On the other hand, to generate $\ket{E,\nu}$ (early with low decoy value), only $D_{e,\ell}^{(1)}$ fires an output. This strategy provides the two electrical pulse amplitudes needed, which can be tuned by changing the comparators supply voltages.\par
    
    The same holds for $\ket{L, \mu}$ and $\ket{L, \nu}$ states, with the only difference that signals are 
    {applied at the time $T_L$, which is shifted from $T_E$} of the temporal distance between 
    the time bin state, that is $10$ns in the case of our experiment. The other two channels are used to control the generation of the $\ket{-,\mu}$ and $\ket{-,\nu}$ states, hence the timing of those pulses 
    {corresponds to $T_-$, i.e. the time in which} the optical pulse travels the phase modulator before entering the 
    {Mach-Zehnder} interferometer. Also in this case the supply voltages are tuned in order to have the same average amount of photons $\mu$ and $\nu$ , equal to those of the early or late state. The electronics operation is summarized in Table \ref{tab:appendix_electrical_signals_table}.\par
    
    \begin{table}
        \centering
        \caption{Signals generated by the FPGA and corresponding analog pulse output for all needed states and decoy levels.}
        \label{tab:appendix_electrical_signals_table}
        \begin{ruledtabular}
            \begin{tabular}{l|l|llll|l|l}
                State & Decoy & \multicolumn{4}{l|}{FPGA outputs} & time slot & $V_O$ \\ \hline
                 & & \multicolumn{1}{c}{$D_{e,\ell}^{(1)}$} & \multicolumn{1}{c}{$D_{e,\ell}^{(2)}$} & \multicolumn{1}{c}{$D_{c}^{(1)}$} & \multicolumn{1}{c|}{$D_{c}^{(2)}$} & & \\ \hline
                $\ket{E}$ & $\mu$ & \multicolumn{1}{c}{1} & \multicolumn{1}{c}{1} & \multicolumn{1}{c}{0} & \multicolumn{1}{c|}{0} & $\Delta T_e$ & $V_{e,\ell}^{(1)} + V_{e,\ell}^{(2)}$ \\ [0.5ex]
                $\ket{E}$ & $\nu$ & \multicolumn{1}{c}{1} & \multicolumn{1}{c}{0} & \multicolumn{1}{c}{0} & \multicolumn{1}{c|}{0} & $\Delta T_e$ & $V_{e,\ell}^{(1)}$ \\ [0.5ex]
                $\ket{L}$ & $\mu$ & \multicolumn{1}{c}{1} & \multicolumn{1}{c}{1} & \multicolumn{1}{c}{0} & \multicolumn{1}{c|}{0} & $\Delta T_{\ell}$ & $V_{e,\ell}^{(1)} + V_{e,\ell}^{(2)}$ \\ [0.5ex]
                $\ket{L}$ & $\nu$ & \multicolumn{1}{c}{1} & \multicolumn{1}{c}{0} & \multicolumn{1}{c}{0} & \multicolumn{1}{c|}{0} & $\Delta T_{\ell}$ & $V_{e,\ell}^{(1)}$ \\ [0.5ex]
                $\ket{-}$ & $\mu$ & \multicolumn{1}{c}{0} & \multicolumn{1}{c}{0} & \multicolumn{1}{c}{1} & \multicolumn{1}{c|}{1} & $\Delta T_c$ & $V_{c}^{(1)} + V_{c}^{(2)}$ \\ [0.5ex]
                $\ket{-}$ & $\nu$ & \multicolumn{1}{c}{0} & \multicolumn{1}{c}{0} & \multicolumn{1}{c}{1} & \multicolumn{1}{c|}{0} & $\Delta T_c$ & $V_{c}^{(1)}$ 
        \end{tabular}
        \end{ruledtabular}
    \end{table}

    For what concern the laser driving electronics, the FPGA board provides a 10MHz square signal, that triggers a pulse generator, whose output is sent to the laser diode module.

\section{Model of imperfections}
\label{appendix:imperfections}
    In the proposed scheme introduction of Sec. \ref{sec:proposed_scheme}, calculations are made assuming all the BSs to be ideal, so having a splitting ratio $T = R = \frac12$. Real devices present natural discrepancies from this value. Here the effect is quantified.\par
    With reference to Figure \ref{fig:sketch_seppia_four_dim}, we label the transmittance of BS $i$ as $T_i$, with the associated reflectance $R_i = 1 - T_i$. Given an input complex amplitude $E_0$ at the input port of $\mathrm{BS}_1$, the amplitudes entering the loop in the  {CW} and  {CCW} modes have the amplitudes of Eq. \ref{eq:amplitudes_just_after_the_input_BS}.
    \begin{subequations}
    \label{eq:amplitudes_just_after_the_input_BS_general} 
        \begin{align}
            E_{CW} &= \sqrt{T_1} E_0
            \label{eq:amplitudes_just_after_the_input_BS_CW} \\
            E_{CCW} &= i\sqrt{R_1}E_0 
            \label{eq:amplitudes_just_after_the_input_BS_CCW}
        \end{align}
    \end{subequations}

    After travelling the loop, the effects of the phase modulators and the unbalanced interferometer leads to amplitudes, of the early ($e$) and late ($\ell$) pulses in the CW and CCW modes, shown in Eq. \ref{eq:amplitudes_before_recombination_at_the_BS_general}.
    \begin{subequations}
        \label{eq:amplitudes_before_recombination_at_the_BS_general}
        \begin{align}
            E_{CW}^{(e)}  &= \sqrt{T_1T_2T_3}E_0 e^{i\phi_c} \\
            E_{CW}^{(\ell)}   &= -\sqrt{T_1R_2R_3} E_0 e^{i\phi_c} e^{i\omega\tau}\\
            E_{CCW}^{(e)} &= i\sqrt{R_1T_2T_3} E_0 e^{i\phi_e} \\
            E_{CCW}^{(\ell)}  &= -i\sqrt{R_1R_2R_3} E_0 e^{i\phi_\ell} e^{i\omega\tau}
        \end{align}
    \end{subequations}

    Recombining at input beamsplitter $\mathrm{BS_1}$, early and late components amplitudes are the one reported in Eq. \ref{eq:early_output_amplitude_general}.
    \begin{subequations}
        \label{eq:early_output_amplitude_general}
        \begin{align}
            E^{(e)}  &= \sqrt{T_1} E_{CW}^{(e)} + i \sqrt{R_1} E_{CCW}^{(e)} \\
                     &= \sqrt{T_2T_3}E_0 (T_1e^{i\phi_c}-R_1e^{i\phi_e})\\
            E^{(\ell)}  &= \sqrt{T_1} E_{CW}^{(\ell)} + i \sqrt{R_1} E_{CCW}^{(\ell)}  \\
                        &=-\sqrt{R_2R_3}E_0e^{i\omega\tau}(T_1e^{i\phi_c}-R_1e^{i\phi_\ell})
        \end{align}
    \end{subequations}

    The corresponding associated transmittances are the in Eq. \ref{eq:transmittance_early_late_general}.
    \begin{subequations}
        \label{eq:transmittance_early_late_general}
        \begin{align}
            T^{(e)} = \frac{|E^{(e)}|^2}{|E_0|^2} = 
            T_2T_3\left(T_1^2 + R_1^2 -2T_1R_1\cos{(\phi_c - \phi_e)}\right) \\
            T^{(\ell)} = \frac{|E^{(\ell)}|^2}{|E_0|^2} = 
            R_2R_3\left(T_1^2 + R_1^2 -2T_1R_1\cos{(\phi_c - \phi_\ell)}\right)\
        \end{align}
    \end{subequations}

    From this it is possible to identify the effect of the non-ideality of the two BSs composing the interferometer as an unbalance in the maximum transmittance of the two time slots. This can introduce an asymmetry when creating states states in the superposition of $\ket{E}$ and $\ket{L}$. The unbalance, if non-negligible, can be corrected by introducing a certain amount of attenuation in one of the two Mach-Zehnder interferometer arms.\par
    
    The performances of the proposed scheme, in terms of fidelity in the prepared state in the $\{\ket{E},\ket{L}\}$ basis, depends instead on the splitting ratio $T_1$ of $\mathrm{BS}_1$. The effect can be quantified by looking at the device transmittance when no modulation is applied. Ideally it is zero, an became nonzero for any value of $T_1 \neq \frac12$. When preparing the $\ket{E}$ or $\ket{L}$ states, the device has to transmit the maximum amount of light in one of the time slots, while completely blocking the light from passing in the other. Assuming here $\mathrm{BS}_2$ and $\mathrm{BS}_3$ to be ideal, the maximum transmittance as function of $T_1$ is expressed by equal to $T_{max} = \frac{1}{4}$ for any value of $T_1$, resulting from the transmittance expression of Eq. \ref{eq:transmittance_early_late_general} putting $\phi_c - \phi_e = \frac{\pi}{2}$. The minimum transmittance, i.e. for $\phi_c - \phi_e = 0$, results instead to be 
    
    \begin{align}
        T_{min} = \frac{1}{4} \left( 2T_1 - 1 \right)^2
    \end{align}
    
    From those, the minimum QBER that can be achieved is 
    \begin{align}
        QBER_{min} = \frac{T_{min}}{T_{min} + T_{max}} = \frac{4T_{min}}{4T_{min} + 1}
    \end{align}

\begin{figure}[h!]
        \centering
        \includegraphics[width=0.8\columnwidth]{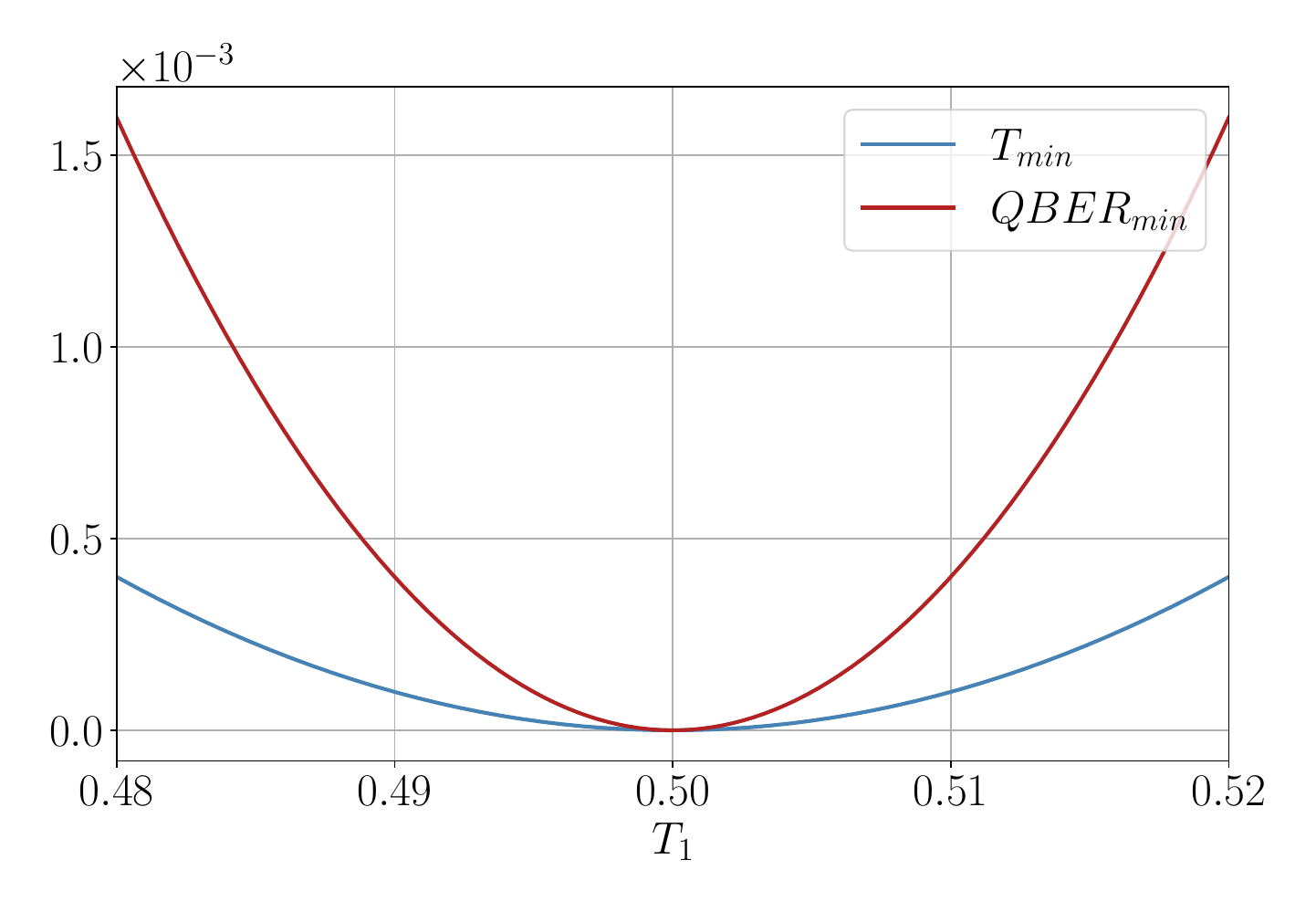}
        \caption{Minimum transmittance and minimum QBER the scheme can achieve as function of the input BS splitting ratio $T_1$.}
        \label{fig:Transmittance_versus_splitting_ratio}
    \end{figure}

    The minimum transmittance $T_{min}$ and the minimum QBER as function of the input BS transmittivity $T_1$ is shown in Figure \ref{fig:Transmittance_versus_splitting_ratio}.

\vfill\eject
\newpage
\bibliography{biblio}
\newpage\hbox{}\thispagestyle{empty}\newpage

\end{document}